\RequirePackage{ifpdf}
\ifpdf % We are running pdfTeX in pdf mode
\documentclass[pdftex]{sigma}
\else
\documentclass{sigma}
\fi

\begin{document}

\allowdisplaybreaks

\renewcommand{\thefootnote}{$\star$}

\renewcommand{\PaperNumber}{080}

\FirstPageHeading

\ShortArticleName{Quantum Integrable  1D anyonic  Models}

\ArticleName{Quantum Integrable 1D anyonic  Models:\\  Construction
  through Braided Yang--Baxter Equation\footnote{This paper is a
contribution to the Proceedings of the International Workshop ``Recent Advances in Quantum Integrable Systems''. The
full collection is available at
\href{http://www.emis.de/journals/SIGMA/RAQIS2010.html}{http://www.emis.de/journals/SIGMA/RAQIS2010.html}}$\!$}

\Author{Anjan KUNDU}

\AuthorNameForHeading{A.~Kundu}

\Address{Theory Group \& CAMCS, Saha Institute of Nuclear Physics, Calcutta, India}
\Email{\href{mailto:anjan.kundu@saha.ac.in}{anjan.kundu@saha.ac.in}}
\URLaddress{\url{http://www.saha.ac.in/theory/anjan.kundu/}}

\ArticleDates{Received May 25, 2010, in f\/inal form October 03, 2010;  Published online October 09, 2010}

\Abstract{Applying braided Yang--Baxter equation
quantum integrable and   Bethe ansatz  solvable 1D anyonic lattice and f\/ield
models are constructed.
  Along with known    models we discover
  novel   lattice  anyonic and $q$-anyonic models as well as
    nonlinear Schr\"odinger  equation
(NLS) and the derivative NLS  quantum  f\/ield  models involving  anyonic
operators,
$N$-particle sectors of which yield the well known    anyon gases,
 interacting through $\delta $ and derivative $ \delta$-function
potentials.}


\Classification{16T25; 20F36; 81R12}

\section{Introduction}

Anyons \cite{wilcz82}  are  receiving renewed  attention
after  their experimental conf\/irmation \cite{anyExp} and the promise of
their   potential
applications in quantum computation \cite{anyKitaev}.
 Although the anyons live
% exchange statistics is  manifested only
 in  two
space-dimensions,  they   remarkably  retain their  basic   properties
  when projected to    one-dimension (1D)
\cite{korany}. Therefore, since
  exactly solvable models are possible to  construct in 1D, interacting
 1D anyonic  models introduced in \cite{kun99,kunbach08,anyspin08}
  are becoming  increasingly popular
in recent years
 \cite{korany,1Dany}, with experimental verif\/ication of 1D anyon
 realised just recently
\cite{1DanyExp}.
However,  though these  models
capture    the exchange algebra of anyonic operators   at the space-separated
points $x \neq y $, they  behave  like bosons
\cite{kun99,kunbach08,korany,1Dany} or fermions  \cite{anyspin08} at the coinciding
points, due to   their inability to   reproduce the required
 anyonic commutation relation (CR) at  $x = y $
   and consequently the corresponding operators    can not
   interpolate  between bosonic and fermionic operators  in
the entire domain. To resolve this problem    in the known
   1D anyonic models
 remains therefore as   a    challenge.

Another  major unsettled  problem   is that, unlike the
   nonlinear Schr\"odinger  equation (NLS) and the derivative
  NLS,  which  give the known  solvable Bose gases
\cite{LiebLin65,Snirman94} at their $N$-particle sectors, no
  anyonic quantum f\/ield models are discovered yet, which could yield
 the known anyon gases at the $N$-particle case,
  \cite{kun99,kunbach08}.

Note however  that the
 anyons,
not commuting at space-separated points,
belong to   nonultralocal models  and go beyond  the standard
formulation of the  quantum inverse scattering method (ISM)   \cite{Fadrev} and
therefore, for constructing    quantum integrable anyonic  models
we have to use the nonultralocal extension of the ISM introduced in
\cite{bybe,nulmkdv}.
Our aim here is to resolve both the above mentioned    problems
in the present 1D anyonic models, by f\/inding
quantum integrable   anyonic lattice and f\/ield models, based on the
braided extension of the Yang--Baxter equation (BYBE)~\cite{bybe,nulmkdv}.
However, we intend to announce here  our preliminary results only,  a~detail
account of which,  focusing on all aspects of the present result as well as our other
related f\/indings, will be given elsewhere in a separate publication~\cite{kundu10}.

The arrangement of the paper is as follows. Section~\ref{section2} describes brief\/ly  the
introduction  of    1D anyonic  models.
 Section~\ref{section3} presents  the BYBE and the related quantum integrable nonultralocal
models.  Subsection~\ref{section3.3} constructs  the quantum integrable anyonic  lattice
  and the   anyonic NLS  f\/ield models.  Subsection~\ref{section3.4} accounts for the
 $q$-anyonic   and the derivative NLS  anyonic f\/ield models.
 Section~\ref{section4} is the     concluding section
followed by the bibliography.

\section{Exactly solvable 1D boson and anyon gases}\label{section2}

Anyons   continue to
 exhibit nontrivial exchange and cross-over properties  even
when projected to  1D,  with
the  2-particle anyonic wave function showing a mixed symmetry
under exchange
\begin{gather}  \Phi (x_{1},x_{2}) =e^{-i \theta}
 \Phi (x_{2},x_{1})
,\label{any1}
\end{gather}
interpolating between  symmetric and antisymmetric wave functions.
 Anyonic wave function also exhibit
 an intriguing sensitivity on the boundary condition:
\begin{gather}
\Phi (x_{1}+L,x_{2}) =  e^{-2i \theta}
    \Phi (x_{1},x_{2}+L), \label{any2}
\end{gather}
on a chain of length $L $ \cite{korany},
 confirming  that the {\it
passing}  of particle~1  through~2 in~1D is not the same as particle~2 {\it
passing} through~1. This ref\/lects the known  property of the standard  2D anyon,
where the ef\/fect of  particle~1  going around~2 is  dif\/ferent  from
that of 2 going around~1. Note that from~\eqref{any1},~\eqref{any2}
 one recovers the usual bosonic behavior at
 $ \theta=0,  $ while   $\theta=\pi $
corresponds to  the   fermion.
The focus on   1D anyon is intensif\/ied in recent years, since
together
with   preserving  the  basic properties of the standard anyon,
  in 1D anyons can   be constructed
as  exactly solvable models  of\/fering    detailed  analytic result, which
should be  valuable for analyzing the standard anyons.

Introduction of  the  exactly solvable anyonic  models in 1D
went through some exciting events with the active involvement of the present
author.
Solvable many particle
bosonic models,  interacting through  singular potentials were known for
long.  The  celebrated model of  Bose gas with   $\delta $-function potential
was  introduced way back in 1963 by Lieb \& Liniger \cite{LiebLin65}.
After about thirty years another
  Bose gas model, interacting through
 derivative $\delta $-function  potential
 was   proposed~\cite{Snirman94}.
Both these bosonic models were solved exactly by the Bethe ansatz.
After such a  success of Bose gases with singular potentials,  there were
naturally   attempts
to build   other interacting bosonic models with
  higher singular potentials,
like   {\it double} $\delta$-function potentials of the form
\begin{gather}
 \gamma_1 \sum_{\langle j,k,l\rangle }
 \delta ({x_j-x_k})
 \delta ({x_l-x_k}) + \gamma_2 \sum_{\langle k,l\rangle}
 (\delta  ({x_k-x_l}))^2,
\label{dbdelta}
\end{gather}
involving 2- and 3-particle  interactions.
 However such attempts  remained  unsuccessful until
the introduction of   $\delta$-function anyon gas by the present author~\cite{kun99},
 which is an  exactly solvable 1D  model and
 shown indeed to be
equivalent to a  double $\delta $-Bose gas involving higher singular
potentials~\eqref{dbdelta} with
its coupling constants  related by $ \gamma_1=\gamma_2=\kappa^2 $.
This anyonic model demonstrated clearly, contrary to the prevailing belief, that
 the the Bethe ansatz is
applicable beyond symmetric or antisymmetric wave functions.

 Another exactly solvable 1D anyon gas model, interacting through
 derivative $\delta $-function potential, is also proposed  recently~\cite{kunbach08},
with the participation of the present author.
When these anyonic gas models, with
dif\/ferent  research groups focusing on their  various aspects,  were
 gai\-ning    popularity~\cite{1Dany}, other
 interesting   nearest neighbor lattice anyonic models,
solvable by the  algebraic Bethe ansatz, were
 proposed  quite recently~\cite{anyspin08}.
However, as mentioned,
    1D anyonic  models proposed so far suf\/fer from the common
  drawback that, they
 behave like bosons or fermions at the coinciding points.

Another important
 well known fact is that, for every
  exactly solvable  Bose gas model, there exists a  quantum integrable
bosonic
f\/ield model,
 $N $-particle sector
of which corresponds to the interacting Bose gas.
In fact the    nonlinear Schr\"odinger equation (NLS)
\begin{gather*}
{H}^{(b1)}
=\int dx \big( \psi^\dag_x \psi_x  + c  (\psi^\dag \psi)^2\big)
%\label{bNLS}
\end{gather*}
 in  bosonic   f\/ield
 $ [\psi (x), \psi^\dag(y)]= \delta(x-y)$ corresponds to
the  $\delta $-Bose gas, while    the quantum
integrable   derivative  NLS f\/ield model yields in the $N $-particle case
the  derivative $\delta $-function Bose gas.
 However an important question surrounding the  1D anyon gases,
 that remained unanswered up to this date
  is
that, what are
the quantum integrable   anyonic f\/ield  models, $N $-particle sectors of which
could generate the known  anyon gas models  interacting through~$\delta $  and
derivative $\delta $-function potentials.  Our aim here is to
take up this challenging  problem and
   discover the needed
   quantum  integrable  anyonic QFT   models.
We also intend to
 construct  integrable anyonic  lattice and f\/ield models
in a systematic way through Yang--Baxter equation, so as to guarantee
 the anyonic commutation relations of its operators   at all points,
 inclu\-ding the coinciding
points, thus  resolving an existing problem  of the anyonic models mentioned
above.\looseness=1

\section[Construction of integrable anyonic models through braided
Yang-Baxter equation]{Construction of integrable anyonic models\\ through braided
Yang--Baxter equation}\label{section3}

Anyonic models due to the  non commutation of anyonic f\/ields  at space
separated points belong to the class of nonultralocal models and
 go beyond the
standard formulation of quantum integrable systems based on the  YBE
\cite{Fadrev}.
We have to
use therefore  an  extension of the YBE with nontrivial braiding
 (BYBE) developed by us \cite{bybe,nulmkdv}, for
systematic generation of the anyonic commutation relations as well as for the construction of
quantum integrable anyonic models.

\subsection[Braided  Yang-Baxter equation]{Braided  Yang--Baxter equation}\label{section3.1}

The BYBE represents two dif\/ferent  commutation relations for the Lax operator $
L_{aj}(u)$,
given at the coinciding and noncoinciding points, expressed through the
standard quantum ${R}(u-v) $-matrix in addition to a braiding matrix $Z $:
\begin{gather}
{R}_{12}(u-v)Z_{21}^{-1}L_{1j}(u) Z_{21} L_{2j}(v)
= Z_{12}^{-1}L_{2j}(v)  Z_{12} L_{1j}(u){R}_{12}(u-v), \label{bybe}
\end{gather}
at the lattice sites $ j=1, 2, \ldots,N $,
together  with  the
 braiding relation (BR):
\begin{gather}
 L_{2 k}(v) Z_{21}^{-1}L_{1 j}(u)
= Z_{21}^{-1}L_{1 j}(u)Z_{21} L_{2 k}(v) Z_{21}^{-1}
\label{br}
\end{gather}
for $k>j$, representing nonultralocality, i.e.\ noncommutativity
at space separated points.
Recall that the quantum $R(u-v) $ matrix is a $4 \times 4 $ matrix
\begin{gather}
\label{R}
R(\lambda)=\left(
\begin{array}{cccc}
 a(\lambda) &&&\\
&  b(\lambda) &  c &  \\
&  c &  b(\lambda)  & \\
&&&  a(\lambda)
 \end{array}\right)
\end{gather}
with rational
\begin{gather}
a(\lambda)=\lambda+\alpha, \qquad b(\lambda)= \lambda, \qquad c=
\alpha  \label{Rrat}
\end{gather}
or  trigonometric
\begin{gather}
a(\lambda)= \sin (\lambda+\alpha), \qquad b(\lambda)=
\sin \lambda, \qquad c=
\sin \alpha  \label{Rtrig}
\end{gather} solutions.
 We  consider both of these forms and show that
they would  generate two dif\/ferent classes of anyonic integrable models.
The braiding matrix $Z$ containing the  anyonic parameter~$\theta $, may be
given in the graded form
\[ Z=\sum_{a,b} e^{i\theta (\hat a \cdot \hat
b)}e_{a,b}\otimes e_{b,a}  , \qquad \hat a=0,1 \ \mbox{denotes
anyonic gradings}, \]
which satisf\/ies all the  relations as required for the braided
generalization~\cite{bybe}.
 For a  $4 \times 4 $ matrix with the  choice  $\hat 1 =0$, $\hat 2=1 $
we get the simplest form
\begin{gather}
Z={\rm diag} \big(1,1,1, e^{i \theta}\big)  \label{Z}
\end{gather}
which we use in constructing all our anyonic models. It is evident that
 for $ \theta=0$: $Z=I$,    BYBE \eqref{bybe} reduces to the
standard  YBE
$ {R}(u-v)L_{1j}(u)  L_{2j}(v)
= L_{2j}(v)   L_{1j}(u){R}(u-v) $,
 while   the BR \eqref{br} recovers the
ultralocal condition
  $[ L_{2 k}(v),L_{1 j}(u)]=0$, $k \neq j  $, related  to the
 bosonic  commutativity.

\subsection{Construction of quantum integrable models}\label{section3.2}

For building  the Hamiltonian of the model we have to
construct conserved quantities by switching over from the local to a
global picture,  by def\/ining the transfer matrix  as a global quantum operator
acting on the multi-particle Hilbert space:
\begin{gather}
\tau(u) = {\rm tr}_a({L}_{a1}(u)\cdots {L}_{aN}(u)) ,\label{tau-C}
\end{gather}
which generates all conserved quantities
 $ \log \tau(u)=\sum_n C_n u^n $.
The BYBE guarantees  that  $[\tau(u),\tau(v)]=0 $, and hence
 the commutativity of the conserved operators $[C_n,C_m]=0 $, ensuring  the
  quantum
integrability of the model. Hamiltonian of the  model  can be chosen
as any of the conserved operators: $H=C_n$, $n=1, 2, 3, \ldots $, which
can therefore be
constructed via~\eqref{tau-C} from
the Lax operator $L_{j}(u) $, satisfying  the BYBE~\eqref{bybe}  with  the known solution of the
quantum  $R $-matrix  and the braiding matrix~$Z $.

\subsection{Rational class of anyonic models}\label{section3.3}

Let us consider f\/irst the rational $ R$-matrix  solution \eqref{R}, \eqref{Rrat}
 together with the
$Z$-matrix~\eqref{Z}, by taking the Lax operator in a general form
 \begin{gather}
  L_{a(l)}^{b}(\lambda)
=\lambda \delta_{ab}   p^{0(l)}_{b}+ \alpha
 p^{(l)}_{ba}
, \label{Lrat}
\end{gather}
 involving abstract anyonic operators
 ${\bf p} \equiv \big(p^{(l)}_{ba},   p^{0(l)}_{b}\big)$, $a,b =1,2$.
The anyonic commutation relations (CR) are obtained directly from the BYBE and
the BR  as
\begin{gather*}
p^{(l)}_{12}    p^{(l)}_{21}-s^{-1} p^{(l)}_{21}    p^{(l)}_{12}=
 p^{(l)}_{11} p^{0(l)}_{2}- s^{-1 }
p^{0(l)}_{1}, \\
 p^{(k)}_{12}    p^{(j)}_{12}= s^{-1  }
 p^{(j)}_{12}    p^{(k)}_{12}, \qquad k>j. %\label{pany}
 \end{gather*}
  Dif\/ferent realizations of these   operators would generate
dif\/ferent  anyonic models with the corresponding  anyonic
algebra,  presented below.  Detailed account of these models
 will be given elsewhere~\cite{kundu10}.

i) {\bf Lattice hard-core  anyonic model.}
 Implementing the above described  scheme for the
 construction
of  lattice anyonic models and realizing
 the general ${\bf p} $  operator as
 \begin{gather*}
 p^{(k)}_{12}=a_k ^\dagger, \qquad     p^{(k)}_{21}= a_k , \qquad
p^{(k)}_{11}= n_k, \qquad p^{(k)}_{22}= 1-n_k, \\
  p^{0(k)}_{1}=1+(s-1)n_k, \qquad n_k= a_k ^\dagger a_k, %\label{pa}
  \end{gather*}
  with an additional hard-core condition $a^2_k=0 $,
we can construct a nearest-neighbor  interacting  anyonic model,
 proposed recently~\cite{anyspin08}:
\begin{gather*} C_1= H^{(1)}
= \sum _{k=1}^N 2 n_kn_{k+1}+
a_ka^\dagger_{k+1}+a^\dagger_k a_{k+1}
, \qquad n_k \equiv a^\dagger_k a_{k}.%\label{NNany}
\end{gather*}
Operators $a^\dagger_k$,  $a_{k} $ obey
 the anyonic CR at space-separated points  $k>l $:
\begin{gather*}
a_k a_l^\dagger=e^{i \theta}
 a_l^\dagger a_k, \qquad a_k a_l=e^{-i \theta} a_l a_k,
%\label{anycrHC}
\end{gather*}
with $\theta=0 $ giving  commuting bosonic and  $ \theta=\pi$ anticommuting  fermionic relations.
  but behave like a fermion  with the anticommutation  relation
$  [a_k , a_k^\dagger]_+=1$,   at the coinciding points,
 conf\/irming the  def\/iciency of the existing anyonic models, mentioned above.

 ii) {\bf Anyonic lattice  model.}  For
 a dif\/ferent realization of the general operator ${\bf p} $ satisfying
 BYBE:
 \begin{gather*}
 p^{(k)}_{12}=  \psi _k, \qquad
    p^{(k)}_{21}=  \tilde \psi_k  ,  \qquad
 p^{(k)}_{11}=n_k \equiv  p_{k}+  \tilde \psi _k
 \psi_k,
%\label{realiz2}
\end{gather*}
together with some other conditions, but without hard-core restriction,
 anyonic operators
$\psi _k$, $\tilde \psi_k$, $p _k $
  can  construct another quantum
 integrable anyonic model
with next-nearest neighbor
interactions  and higher order nonlinearity given by the Hamiltonian
\begin{gather}
 C_3= {\rm H}^{(2)}= \sum_k \left(\tilde \psi_{k+1}
\psi_{k-1}  -(n_{k}+
n_{k+1})\tilde \psi_{k+1}
\psi_{k} + \frac 1{3 \Delta ^2}  n^3_k  \right)
\label{anyLNLS}
\end{gather}
Remarkably,  this model  gives
 the needed anyonic  CR  at the    coinciding points $k$:
\begin{gather}
 \psi_{k} \tilde \psi_{k} -
 e^{-i \theta} \tilde \psi_{k}  \psi_{k} =   p_{k}
\label{anyCRx0}
\end{gather}
together with
\begin{gather}
 \psi_{k} \tilde \psi _{j}=e^{i \theta}
\tilde \psi _{j} \psi_{k} \label{anyCRx+}
\end{gather}
and other similar relations at noncoinciding  points  $k>j $.
 Thus  we  solve here
one of the  existing problems of
 anyonic models by constructing
 the anyonic operator with  proper CR, valid  at all points.

  iii) {\bf Quantum integrable   NLS  anyonic f\/ield model.}
Taking carefully the continuum limit of the
lattice anyonic model \eqref{anyLNLS} with
  $ k \to x $, and the f\/ield $
\psi_k \to A(x)$, $\tilde \psi_k \to A^{\dagger}(x) $, we can derive a
 novel  NLS anyonic f\/ield model
 \begin{gather}
\hat H^{(3)}= \int dx \big(A^\dagger_{x}A_x+c (A^\dagger A)^2 \big) \label{anyLNS}
 \end{gather}
with the  anyonic f\/ield operator $A(x) $  obeying
the needed  CR at all points, obtained from its lattice counterpart
\eqref{anyCRx0}, \eqref{anyCRx+}. At the coinciding points $x\to y^+ $
we get  the anyonic CR for the f\/ield operators as
\begin{gather*}
A(x) A^\dagger(y)-e^{i \theta}A^\dagger(y)A(x)=\delta(x-y)  %\label{anyx0}
\end{gather*}
together with the anyonic relations  at $x>y $:
\begin{gather*}
 A(x) A^\dagger(y)=e^{i \theta} A^\dagger(y) A(x),
\qquad
A(x) A(y)=e^{-i \theta} A(y) A(x).   %\label{anyx2+}
\end{gather*}
Clearly these anyonic operator relations can interpolate between
 the bosonic  (at $\theta=0$) and  fermionic (at $\theta=\pi$)
 CR at all points.

We solve another outstanding problem of the  anyonic
models
by f\/inding the
$N$-particle sector
\begin{gather*}
|N\rangle =\int d^N x \sum_{\{x_l\}} \Phi (x_1,x_2,\ldots, x_N)
   A^\dagger (x_1)A^\dagger (x_2)\cdots   A^\dagger (x_N)  |0\rangle  %\label{Nsector}
\end{gather*}
  of the NLS anyonic f\/ield  model  \eqref{anyLNS}, which   indeed gives
the well known  $\delta $ -function  anyon gas model
\[
H_N= -\sum_k \partial^2_k +c \sum_{k \neq j} \delta (x_k -x_l)
 \]
 linking it thus with a genuine 1D anyonic quantum  f\/ield model.
All the above anyonic models are obtained using the rational $R$-matrix
solution. Now we switch over to the trigonometric case, presenting only our main
result, with all details intended for a separate report~\cite{kundu10}.

\subsection{Trigonometric class of anyonic models}\label{section3.4}

 We consider now   the trigonometric quantum  $R$-matrix \eqref{R}, \eqref{Rtrig}
 related to the $ xxz$ spin-$\frac 1 2 $ chain and a corresponding  trigonometric
modif\/ication of the Lax operator \eqref{Lrat} through abstract
anyonic $q$-oscillator operator ${\boldsymbol \tau} \equiv (\tau_{12},
\tau_{21}, \tau^{\pm}_a,\; a=1,2 ) $, keeping  however  the same
 braiding
 mat\-rix~$Z $~\eqref{Z}. Following the scheme for constructing the integrable
 models starting from the Lax operator as discussed above and using
 dif\/ferent concrete realizations of the operator ${\boldsymbol \tau} $,    we  gene\-ra\-te
 a  trigonometric
class of anyonic lattice and f\/ield
 models, with a deformation parameter  $q=e^{i \alpha} $, in addition to the
anyonic parameter~$\theta $. Dif\/ferent forms of  $q$-anyonic algebras
can be derived from the BYBE and the BR relations for
   dif\/ferent realizations of  $ {\boldsymbol\tau}$.

iv) {\bf  $\boldsymbol{q}$-anyonic model.}
Using a realization through quantum anyonic operators:
\begin{gather*}
\tau^{\pm}_1=q^{\mp N}, \qquad \tau^{\pm}_2= q^{\pm N}e^{i \theta N},
\qquad \tau_{12}=
\tilde \phi, \qquad \tau_{21}=
\phi,   %\label{tau-phi}
\end{gather*}
 we get  from   BYBE \eqref{bybe}  and \eqref{br}
an
{\it  anyonic $q$-oscillator} model with CR:
\begin{gather*}
\phi_{k} \tilde \phi_{k} -
 e^{i \theta}   \tilde \phi_{k}  \phi_{k} =  e^{i \theta N_k}
 \cos 2 \alpha N_k %\label{qanyx0}
 \end{gather*}
at the coinciding points and
\begin{gather*}
\phi_{k} \tilde \phi_{j} =
e^{i \theta }
\tilde \phi_{j} \phi_{k}, %\label{qanyx+}
\end{gather*}
 at the space-separated points  $ k >j$, showing  that
 we can generate through our scheme the anyonic algebra relations
both at the coinciding and noncoinciding points, even for the  quantum
deformation, which was another aim we  started with.
 Now we intend to f\/ind the corresponding integrable quantum f\/ield model,
which
is our other major aim.
Therefore relegating   further details of this
 anyonic $q$-oscillator model~\cite{kundu10},
we   go to the continuum limit of this model and
 present below another
anyonic quantum integrable f\/ield model.

  v) {\bf Derivative NLS anyonic f\/ield model.}
At the  f\/ield limit by taking  lattice constant  $\Delta \to 0$,
 the anyonic $q$-oscillator would reduce to an anyonic f\/ield operator:
$\phi_k \to D(x) $,
to derive from   the $q$-anyonic lattice model   a
 quantum integrable derivative NLS  anyonic f\/ield model
\begin{gather*}
\hat H^{(5)}= \int dx
\big(D^\dagger_{x}D_x+2i \kappa  (D^\dagger )^2 D D_x\big) %\label{anyDNLS}
\end{gather*}
with the anyonic f\/ield operator satisfying at $x \to y^+ $  the  CR
\begin{gather}
D(x) D^\dagger(y)-e^{i \theta}D^\dagger(y)D(x)=\kappa \delta(x-y)
 \label{anyCRx0dnls}
\end{gather}
and
\begin{gather}
D(x) D^\dagger(y)=e^{i \theta} D^\dagger(y) D(x),
 \label{anyCRx+dnls}
 \end{gather}
 at $x > y $. Interestingly, even at the f\/ield limit the important anyonic
operator relations remain valid  at the coinciding \eqref{anyCRx0dnls}
 as well as at the  noncoinciding~\eqref{anyCRx+dnls}  points.
  We f\/ind  that, this anyonic DNLS f\/ield model
  gives  at its $N$-particle sector $|N\rangle  $, the
recently proposed  derivative  $\delta $-function  anyon gas model
\[
H^d_N= -\sum_k \partial^2_k +i \kappa
 \sum_{k \neq j} \delta (x_k -x_l) (\partial_{x_k} + \partial_{x_l}),
 \]
establishing thus the  missing link of this  anyon gas
  to a   quantum integrable   anyonic f\/ield model,  discovering  which was
our f\/inal aim.

We have constructed
here   from  the BYBE
a series of anyonic and $q$-anyonic models, namely
i)~nearest-neighbor  hard-core anyonic, ii)~next-nearest-neighbor
 higher nonlinear lattice anyonic and  iii)~quantum NLS anyonic  f\/ield models
from the rational class and  iv)~anyonic $q$-oscillator and  v)~anyonic DNLS quantum f\/ield models from the
trigonometric class. We emphasize that, except the f\/irst one i)~all other  anyonic
models are new, presented here for the f\/irst time and
   are
 quantum integrable models,   solvable by
the algebraic Bethe ansatz. Postponing the relevant details of
 the Bethe ansatz formulation to~\cite{kundu10}  we just mention  that, it  goes
   pretty close to the standard   formulation   for the ultralocal models~\cite{korbook}. The anyonic contribution  entering through the
braiding matrix~$Z$,  due to its simple form considered here,
 does not create much dif\/f\/iculty and follow the line of~\cite{nulmkdv}, though
individual  nonultralocal model presented here  would  have its  specif\/ic problem and solution.

 \section{Concluding remarks}\label{section4}

We have constructed two classes of 1D anyonic
models, rational and trigonometric, in a systematic way starting from the
braided YBE. The known as well as new anyonic models, that we have
constructed, are all quantum integrable and exactly solvable by the algebraic
Bethe Ansatz. Among the new models a next-nearest neighbor interacting
lattice anyonic model and a nonlinear Schr\"odinger anyonic quantum f\/ield
model belong to the rational class. Among the models  belonging to the
trigonometric class we have discovered an anyonic $q$-oscillator lattice model
and a
derivative NLS anyonic f\/ield model.

Remarkably, the NLS  anyonic quantum f\/ield
model  at its $N $-particle sector recovers  the  $\delta $-function  anyon
gas, while  the derivative NLS anyonic f\/ield model is
linked to the recently proposed derivative $\delta $ anyon gas. Thus we f\/ind
the   quantum integrable f\/ield models corresponding to
  the known anyon gas models,   solving  an outstanding problem.

The   anyonic  models we have constructing  here  exhibit proper anyonic CR
 at all points, resolving  another problem  of the existing models.

Investigating further   the  nonultralocal  BYBE in the trigonometric case,
we have  obtained  recently    a
 new kind of  anyonic quantum group Hopf algebra,
with an additional  deformation given by the   anyonic parameter. We hope
that, this
two-parameter quantum
algebra with explicit nonultralocal nature, which is still under study and
will be reported in time \cite{kundu10}, would be a signif\/icant addition in
the knowledge of quantum algebras.

Another promising line of research would be to f\/ind nonabelian realizations
of the integrable 1D anyonic models exploiting the BYBE, which might shed
light to  the  nonabelian anyonic models, importance of which  is growing in
recent years~\cite{anyKitaev}.

\pdfbookmark[1]{References}{ref}

\end{document}